# Thick Accretion Disk and Its Super Eddington Luminosity around a Spinning Black Hole

Uicheol Jang[1], Hongsu Kim[2], Yu Yi[1†]

[1]Department of Astronomy, Space Science and Geology, Chungnam National University, Daejeon 34134, Korea
[2]Center for Theoretical Astronomy, Korea Astronomy and Space Science Institute, Daejeon 34055, Korea

In the general accretion disk model theory, the accretion disk surrounding an astronomical object comprises fluid rings obeying Keplerian motion. However, we should consider relativistic and rotational effects as we close in toward the center of accretion disk surrounding spinning compact massive objects such as a black hole or a neutron star. In this study, we explore the geometry of the inner portion of the accretion disk in the context of Mukhopadhyay's pseudo-Newtonian potential approximation for the full general relativity theory. We found that the shape of the accretion disk "puffs up" or becomes thicker and the luminosity of the disk could exceed the Eddington luminosity near the surface of the compact spinning black hole.

## 1. INTRODUCTION

The inflow of dust and gas into compact objects, such as neutron stars and black holes, is known as accretion flow. Specifically, as dust and gas flow into these compact objects, their temperature rises due to the friction associated with viscosity, resulting in black body radiation. This black body radiation generates electromagnetic waves leading to the illumination of compact objects, which are, by definition, non-luminous. Similarly, non-luminous compact objects or other weakly illuminating compact objects can be observed. The dominant electromagnetic radiation comprises X-rays, with the X-ray binary as the most commonly recognized example. Thus, to fully understand the observational data of non-luminous or weakly illuminating objects, the accretion process needs to be studied in detail. In general, among main sequence stars, massive stars (core mass $M \simeq 3M_\odot$) evolve into black holes, resulting in a strong magnetic field as a consequence of collapse. This preserves the magnetic flux of the progenitor star, which shrinks during the collapse process. The neutron star, which is the endpoint of a supernova explosion, is one such example ($B \sim 10^{12}$ G). This indicates that the formation of accretion disk may involve an electromagnetic effect in addition to the gravitational and hydrodynamic effects. Hence, in addition to gravity, the electromagnetic field and hydrodynamics play a major role in the accretion process. The black hole magnetosphere can be defined in the following manner: inside the surface of the magnetosphere, the magnetic pressure is greater than the gas pressure, contrary to the outside of the magnetosphere surface. In the present work, we explore the effect of an electromagnetic field on the accretion flow, in addition to the effects of gravity and hydrodynamics. The standard model for the accretion disk is the Shakura & Sunyaev model (1973), which assumes a thin accretion disk. However, as we approach a black hole, Shakura-Sunyaev's thin disk model may fail, as a thorough study of the shape of the gravitational potential and the self-radiation pressure of an accretion disk reveals a region where the accretion inflow exceeds the outflow toward the black hole. Consequently, in this region, the accretion disk "puffs up" remarkably and deviates from the standard Shakura-Sunyaev model. Paczyńsky & Witta (1980) studied this phenomenon for the non-rotating black hole case using the pseudo-Newtonian potential for a Schwarzschild black hole. However, in space, all compact objects are spinning, including black holes. Thus, the original work of Paczynsky and Witta has been revised for the case of spinning Kerr black holes that possess mass and angular momentum. The pseudo-Newtonian potential for the spinning Kerr black hole was developed by Mukhopadhyay (2002). In this study, we explore the swollen portion of the accretion disk around a spinning Kerr black hole by following the approach undertaken by Paczynsky and Witta employing Mukhopadhyay's pseudo-Newtonian potential instead of Paczynsky and Witta's pseudo-Newtonian potential, which is used for the non-spinning Schwarzschild black hole case.

## 2. METHODS

### 2.1 Pseudo-Newtonian Potential

To understand the dynamics around black holes, it is essential to employ the general theory of relativity. When dealing with a number of objects, rigorous general relativistic treatment demands huge computing power and is therefore impractical. Hence, in the present work, we employ the pseudo-Newtonian potential, which involves relativistic effects on the Newtonian gravitational potential. For the case of non-spinning black holes, the pseudo-Newtonian potential is known as the Paczynsky–Witta potential, as given by Eq. (1):

$$V(r) = -\frac{GM}{r - 2GM}. \qquad (1)$$

In the case of spinning black holes, the pseudo-Newtonian potential for a spinning uncharged Schwarzschild black hole, Kerr black hole, charged Reissner–Nordstrom black hole, and Kerr–Newman black hole, is known as the Mukhopadhyay potential (Mukhopadhyay 2002), as given by Eq. (2):

$$\begin{aligned}V_x = &\frac{a^2}{2x^2} + \frac{4a}{\sqrt{x}} + \frac{2\left(9a^3 x - 10ax + 13\sqrt{x} - 13a^{2\sqrt{x}} + 6a^3 - 8a\right)}{(27a^2 - 32)\left(x^{3/2} - 2\sqrt{x} + a\right)} \\ &- 2\log x + \frac{2}{27a^2 - 32}\sum_{y = x_1, x_2, x_3}\frac{1}{3y^2 - 2}\log\left(\sqrt{x} - y\right) \\ &\left(54a^2 y^2 - 64 y^2 + 63a^3 y - 75ay - 107^2 + 128\right),\end{aligned} \qquad (2)$$

where

$$x_1 = \frac{2^{4/3}}{p} + \frac{p}{2^{1/3}3},\quad x_2 = -\left(\frac{2^{1/3}q}{p} + \frac{pq^*}{2^{1/3}6}\right),\quad x_3 = x_2^*,$$

$$p = \left(\sqrt{729a^2 - 864} - 27a\right)^{1/3},\text{ and } q = \left(1 + i\sqrt{3}\right).$$

In this equation, a = 0 or a = 1, where a = 0 corresponds to the Paczynsky–Witta potential.

### 2.2 Vertical Structure of the Thick Portion of the Accretion Disk

Owing to the frictional force resulting from viscosity, the differential rotation of layers in the inner portion of the accretion disk produces outward radiation. Consequently, the gas and dust particles falling into a black hole experience outward radiation, which resists the gravitational pull of the black hole, leading to swelling of the accretion disk in the inner region. As shown in Fig. 1, this bulged portion differs substantially from the rest of the accretion disk, which is thin. According to Paczynsky and Witta, the pseudo-Newtonian potential for a Schwarzschild black hole is given by $V(r) = -\frac{GM}{r - r_g}$, where $r_g$ = 2GM, assuming that the thickness of the disk is negligible compared to the radius of the disk. This pseudo-Newtonian potential can be safely approximated using Eq. (3):

$$V_0(r) = V\left(\left(r^2 + z_0^2\right)^{1/2}\right). \qquad (3)$$

To make our computation tractable, we introduce $R \equiv (r^2 + z^2)^{1/2}$, where the effective potential involving the presence of centrifugal force is given by Eq. (4):

$$V_{eff}\vec{r} = V_0(r) - \int_r^\infty r'\left[\omega(r')\right]^2 dr'. \qquad (4)$$

From this, we derive the gravitational acceleration (force per unit mass) as $\overrightarrow{g_{eff}} = -\nabla V_{eff}$. We now assume that the thick portion of the accretion disk is in the equilibrium state to maintain its shape. This assumption then demands that the surface of the accretion disk be on the equipotential plane, which is orthogonal to the direction of gravitational acceleration. This can be represented by the following equation:

$$0 = dV_{eff} = d\left[V_0 - \int r\omega^2 dr\right] = \left(\frac{dV_0}{dr} r\omega^2\right) + \left(\frac{dV_0}{dz}\right) dz$$

$$\therefore \frac{dz}{dr} = \left(r\omega - \frac{dV_0}{dr}\right)\left(\frac{dV_0}{dz}\right)^{-1}. \quad (5)$$

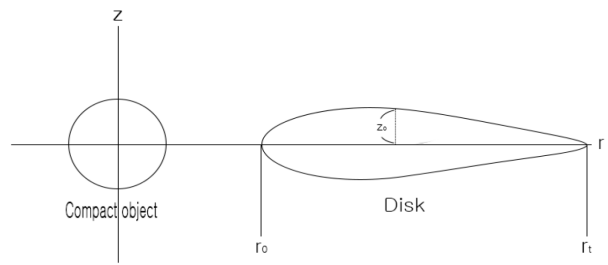

**Fig. 1.** Profile of the thick accretion disk.

From Eq. (5), we obtain Eq. (6).

$$\frac{dV_0}{dr} = \frac{GM}{(R-r_g)^2} \cdot \frac{r}{R}, \quad \frac{dV_0}{dz} = \frac{GM}{(R-r_g)^2} \cdot \frac{z}{R}, \quad (6)$$

where $R \equiv (r^2 + z^2)^{1/2}$.

Using Eq. (6), Eq. (5) can be rewritten as follows:

$$\frac{dz}{dr} = \frac{r}{z}\left[\frac{\omega}{GM} R(R-r_g)^2 - 1\right]. \quad (7)$$

To solve this differential equation, Eq. (7) is represented in a closed form.

$$\frac{2zdz}{2rdr} + 1 = \frac{\omega^2}{GM} R(R-r_g)^2$$

$$\therefore \frac{dR^2}{dr^2} = \frac{\omega^2}{GM} R(R-r_g)^2, \quad (8)$$

where ω is the angular momentum per unit mass (specific angular momentum), and Eq. (8) represents the thickness of the bulging part of the accretion disk. For this calculation, caution should be exercised as ω may not denote the angular frequency for Keplerian motion, unlike the ordinary thin accretion disk. Thus, ω may deviate from the simple Keplerian angular velocity. Notably, until this point, we have considered the case of non-rotating Schwarzschild black holes, and our interest is in the case of rotating Kerr black holes. The rotating Kerr black hole adaptation of the pseudo-Newtonian potential was developed by Mukhopadhyay (2002). Therefore, in the context of the Mukopadhyay pseudo-Newtonian potential, we have

$$\frac{dV_0}{dr} = \frac{1}{G^2 M^2} \frac{r}{X}\left(\frac{dV_0}{dX}\right), \quad \frac{dV_0}{dz} = \frac{1}{G^2 M^2} \frac{z}{X}\left(\frac{dV_0}{dX}\right). \quad (9)$$

By substituting these two equations into Eq. (4), we obtain Eq. (10):

$$\frac{dz}{dr} = \left(r\omega^2 - \frac{dV_0}{dr}\right)\left(\frac{dV_0}{dz}\right)^{-1}$$

$$= \frac{r}{z}\left[G^2 M^2 \omega^2 X \left(\frac{dV_0}{dX}\right)^{-1} - 1\right]\frac{2zdz}{2rdr} + 1 \quad (10)$$

$$= G^2 M^2 \omega^2 X \left(\frac{dV_0}{dX}\right)^{-1}$$

$$\therefore \frac{dX^2}{dr^2} = \omega^2 X \left(\frac{dV_0}{dX}\right)^{-1}.$$

Now, we are able to describe the thick portion of an accretion disk around a black hole.

**2.3 Radiation and Luminosity of the Thick Portion of the Accretion Disk**

According to the Paczynsky–Witta model, in principle, the power per unit area, $F_{rad}$, is given as follows:

$$F_{rad} = \frac{c}{\kappa} g_{eff}, \quad (11)$$

where $g_{eff}$ denotes the acceleration, $c$ the speed of light, and $\kappa$ denotes the opacity per unit mass in the CGS (Gaussian) unit. Next, for the integration to compute the luminosity of the thick portion of an accretion disk, the surface element (dσ) is given by Eq. (12):

$$\cos\theta = \frac{dr}{\sqrt{dr^2 + dz^2}} = \left[1 + \left(\frac{dz}{dr}\right)^2\right]^{-1/2}. \quad (12)$$

In the cylindrical coordinate system,

$$d\sigma (rd\Phi)(\sec\theta dr) = r\left[1 + \left(\frac{dz}{dr}\right)^2\right]^{1/2} d\Phi dr, \quad (13)$$

where $g_{eff} = |-\nabla V_{eff}|$, the vertical acceleration is $g_z = \partial V_0/\partial z$, and $g_{eff} \cos\theta = g_z$. In addition, using the expressions for $F_{eff}$ and $\cos\theta$, $F_{rad}$ is given by Eq. (14):

$$F_{rad} = \frac{c}{\kappa} \frac{\partial V_0}{\partial z} \left[ 1 + \left( \frac{dz}{dr} \right)^2 \right]^{1/2}. \quad (14)$$

Thus, the luminosity of the thick portion of an accretion disk can be represented as follows:

$$F_{rad} = \frac{c}{\kappa} \frac{\partial V_0}{\partial z} \left[ 1 + \left( \frac{dz}{dr} \right)^2 \right]^{1/2} = \frac{4\pi c}{\kappa} \int_{r_0}^{r_t} r \frac{\partial V_0}{\partial z} \left[ 1 + \left( \frac{dz}{dr} \right)^2 \right] dz. \quad (15)$$

Finally, by substituting $dz/dr$ and $\partial V_0/\partial z$ into this equation, we obtain,

$$L_{rad} = \frac{4\pi cGM}{\kappa} \int_{r_0}^{r_t} \left[ \frac{l^4 R}{G^2 M^2 z r^5} (R - r_g)^2 - \frac{2l^2}{GMzr} + \frac{rR}{z(R - r_g^2)} \right] dr \quad (16)$$
$$= L_{Edd} \cdot I_s(r_0, r_t).$$

Here, $L_{EDD} = 4\pi cGM/\kappa = 1.3 \times 10^{38}$ (M/M☉)erg/s is the Eddington luminosity, which is the critical value for the luminosity that an astrophysical object with mass M can radiate. Thus far, we have considered the case of a non-rotating Schwarzschild black hole. Similarly, we can derive the luminosity for a rotating Kerr black hole, as shown below.

$$L_{rad} = \frac{4\pi cGM}{\kappa} \int_{r_0}^{r_t} \left[ \frac{1}{G^2 M^3} \frac{rz}{X} \left( \frac{dV_0}{dx} \right) \right]$$
$$\{1 + \frac{r^2}{z^2} \left[ G^2 M^2 \frac{l^2}{r^4} \left( \frac{dV_0}{dX} \right)^{-1} - 1 \right]^2 \} dr = L_{Edd} \cdot I_k(r_0, r_t), \quad (17)$$

where $I_k$ can be greater than one.

## 3. RESULTS

### 3.1 Analysis of Luminosity for an Accretion Disk

The luminosity of the accretion disk is given by Eq. (17). The relations amount, $X$, $V_0$ and $r$ are determined by the accretion rate, $\dot{M}$, according to the Shakura-Sunyaev model (Shakura & Sunyaev 1967), given by $\dot{M} = \frac{\Delta}{d} \dot{M}_{Edd}$. In general, when the disk is very thin ($\Delta \ll d$), the accretion rate is considerably lower than the Eddington accretion rate. In the present work, we study the thick portion of the disk, where the accretion rate could reach Eddington's critical value. Notably, although an astrophysical object with super Eddington luminosity will blow up and is very unstable, when the accretion flow falls into the central black hole and the disk itself is very soft and flexible, it sustains its structure even at the super Eddington luminosity.

Therefore, we suggest that the extraordinary luminosity of a discovered quasar might be attributable to the "thick" and narrow inner part of the accretion disk surrounding the supermassive black hole, which excluding strong lensing, is the power engine for bright quasars.

### 3.2 Vertical Structure of the Thick Accretion Disk

First, we reproduced the results of the work by Paczynsky–Witta to study the vertical structure of a thin accretion disk in the absence of rotation. The vertical structure has already been characterized in their work, as in Eq. (18).

$$\therefore R = (r^2 + z_0^2)^{1/2} = \frac{r_0 - r_g}{1 - \frac{r_0 - r_g}{GM} \int_{r_0}^{r} r\omega^2 dr} + r_g$$

$$\therefore z_0(r) = \left\{ \left[ \frac{r_0 - r_g}{1 - \frac{r_0 - r_g}{GM} \int_{r_0}^{r} r^{-3/2} dr} \right]^2 - r^2 \right\}^{1/2}, \quad (18)$$

where $l$ denotes the specific angular momentum, which can be written as $\frac{r^{3/2}}{r_0 - 1} = A_\beta (r - r_0)^B$ for numerical analysis. Employing Mathematica, we can plot this specific angular momentum, as shown in Fig. 2, even though the values for $r$ are the same as in the work of Paczynsky–Witta.

The plots above are essentially the same as those in the work of Pazcynsky and Witta. In Figs. 2–4, $r_t$ denotes the transition radius, which specifies the boundary that separates the thin and the thick accretion disk, with $l(r_t) = \frac{r^{3/2}}{r_0 - 1} = A_\beta (r - r_0)^B$. Close inspection reveals that the closer the starting point ($r_0$) is to the compact object, the thicker the accretion disk becomes, and the more the disk spreads out. This allows us to speculate the appearance of accretion disk and ultimately to compare it with the theoretical computations.

The pseudo-Newtonian potential for a spinning object is considerably more complex than that of a non-spinning object. Therefore, to investigate the vertical structure of a thick accretion disk, we adopted a numerical integration. As a result, Fig. 5–Fig. 7 ware obtained. Upon plotting the thick portion of the spinning black hole case, the shape of the disk evidently becomes steeper, particularly for the case where $r_0 = 3 r_g$, where the disk becomes thinner but wider. That is, it is smaller vertically but larger horizontally. We conjecture that this can be attributed to the large centrifugal force of a black hole or a neutron star.

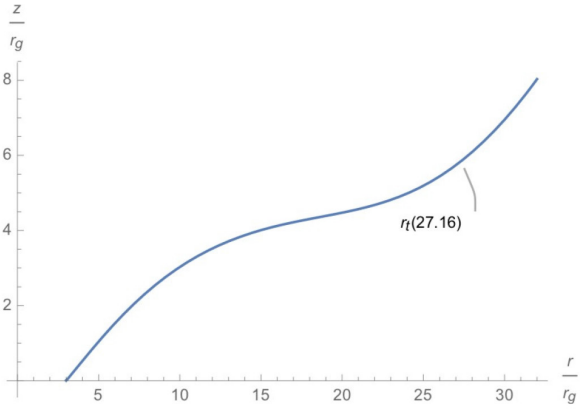

**Fig. 2.** Accretion disk 1 ($r_0 = 3\ r_g$) for Schwarzschild black hole, where $r_t$ denotes the boundary between the thick and the thin disk.

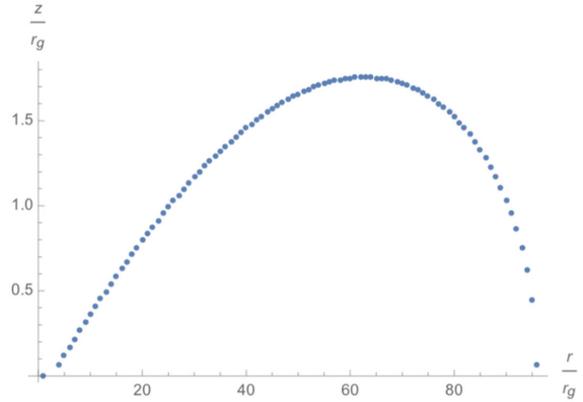

**Fig. 5.** Accretion disk 1 ($r_0 = 3\ r_g$) for a Kerr black hole.

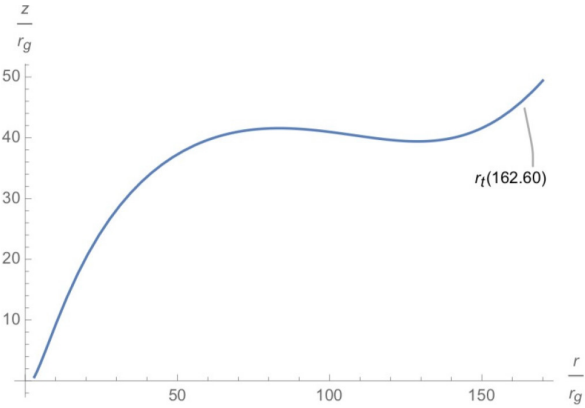

**Fig. 3.** Accretion disk 1 ($r_0 = 2.3\ r_g$) for Schwarzschild black hole, where $r_t$ denotes the boundary between the thick and the thin disk.

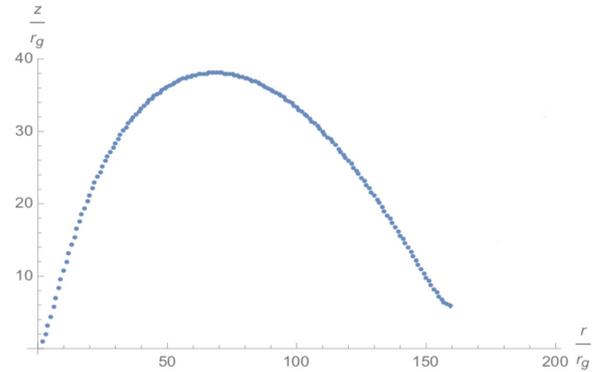

**Fig. 6.** Accretion disk 1 ($r_0 = 2.3\ r_g$) for a Kerr black hole.

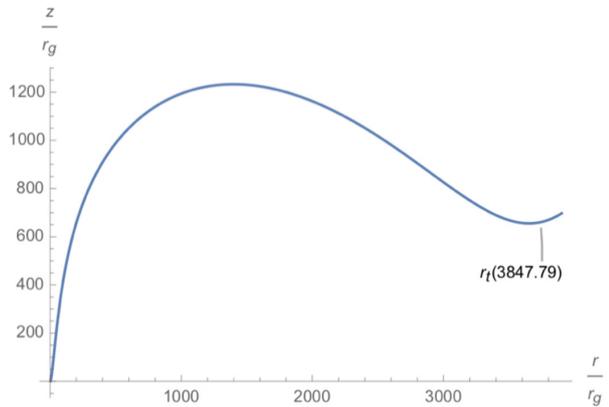

**Fig. 4.** Accretion disk 1 ($r_0 = 2.05\ r_g$) for Schwarzschild black hole, where $r_t$ denotes the boundary between the thick and the thin disk.

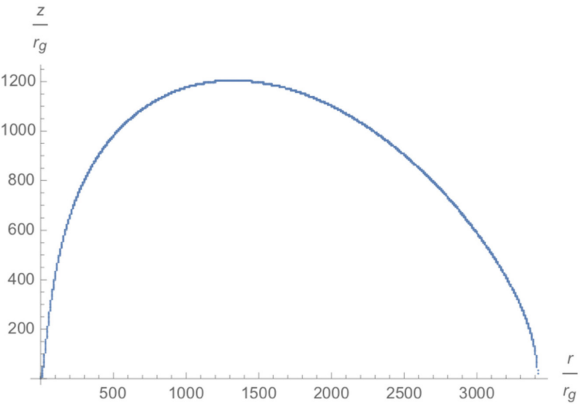

**Fig. 7.** Accretion disk 1 ($r_0 = 2.05\ r_g$) for a Kerr black hole.

## 4. CONCLUSIONS

In the present work, we study the shape and luminosity of a thick accretion disk in conjunction with the gravitational attraction toward the gas and dust inflow. In particular, we investigated the innermost part of an accretion disk that puffs up (Fig. 8); primarily due to the high radiation pressure that resists the accretion inflow in a different manner to that for a thin accretion disk (Fig. 9).

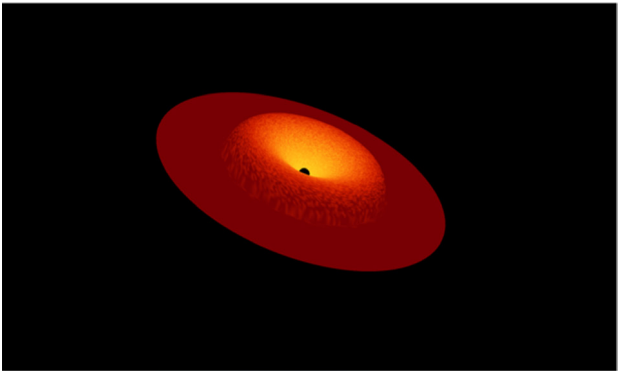

**Fig. 8.** Conceptual picture of the thick accretion disk.

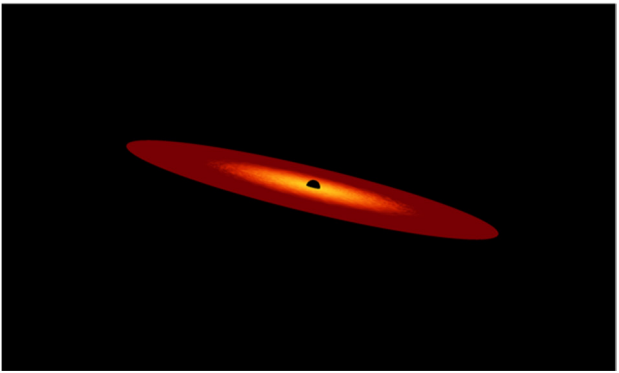

**Fig. 9.** Conceptual picture of the thin accretion disk.

Briefly, we have followed the methodology first employed by Paczynsky and Witta for the case of a non-rotating black hole, employing a different pseudo-Newtonian potential, which was first suggested by Mukhopadhyay for the spinning Kerr black hole case. Owing to the substantially more complex spinning structure, we rely on a numerical integration strategy. In summary, the structure of the thick portion of an accretion disk for a spinning case is wider horizontally than in the non-spinning case. This result implies that centrifugal force plays a role when a centrally compact object rotates. In addition, we can infer that the disk could radiate with a super Eddington luminosity. This model can therefore be a candidate to explain excessively bright quasar observations, such as those listed in Table 1.

**Table 1.** List of bright quasars (brighter than their Eddington luminosity), where $L_\odot$ is solar luminosity

| Quasars | Brightness (Bolometric luminosity) |
|---|---|
| J043947.08 + 163415.7 (Fan et al. 2019) | $5.85 \times 10^{14}\ L_\odot$ |
| SDSS J0100 + 2922 (Wu et al. 2015) | $4.29 \times 10^{14}\ L_\odot$ |
| ULAS J1120 + 0641 (Mortlock et al. 2011) | $6.3 \times 10^{13}\ L_\odot$ |
| ULAS J1342 + 0928 (Bañados et al. 2018) | $4 \times 10^{13}\ L_\odot$ |

## ACKNOWLEDGMENTS

This work was supported by research fund of Chungnam National University.

## ORCID

Uicheol Jang   https://orcid.org/0000-0001-7859-8581
Hongsu Kim    https://orcid.org/0000-0002-1508-3166
Yu Yi              https://orcid.org/0000-0001-9348-454X

## REFERENCES

Bañados E, Venemans BP, Mazzucchelli C, Farina EP, Walter F, et al., An 800-million-solar-mass black hole in a significantly neutral universe at a redshift 7.5, Nature. 553, 473-476 (2018). https://doi.org/10.1038/nature25180

Fan X, Wang F, Yang J, Keeton CR, Yue M, et al., The discovery of a gravitationally lensed Quasar at z=6.51, Astrophys. J. Lett. 870, L11 (2019). https://doi.org/10.3847/2041-8213/aaeffe

Mortlock DJ, Warren SJ, Venemans BP, Patel M, Hewett PC, et al., A luminous quasar at a redshift of z = 7.085, Nature. 474, 616-619 (2011). https://doi.org/10.1038/nature10159

Mukhopadhyay B, Description of pseudo-newtonian potential for the relativistic accretion disks around Kerr black holes, Astrophys. J. 581, 427-430 (2002). https://doi.org/10.1086/344227

Paczyński B, Witta, PJ, Thick accretion disks and supercritical luminosities, Astron. Astrophys. 88, 23-31 (1980).

Shakura NI, Sunyaev RA, Black holes in binary systems. Observational appearance, Astron. Astrophys. 24, 337-355 (1973).

Wu XB, Wang F, Fan X, Yi W, Zuo W, et al., An ultraluminous quasar with a twelve-billion-solar-mass black hole at redshift 6.30, Nature. 518, 512-515 (2015). https://doi.org/10.1038/nature14241